\documentclass[a4paper,11pt]{article}  
\usepackage[toc,title]{appendix}
\pdfoutput=1

\usepackage{graphicx} 
\usepackage{bbold}
\usepackage{mathtools}
\usepackage{amsmath}
\numberwithin{equation}{section} %equation numbering
\usepackage{amsfonts}
\usepackage{physics}
\usepackage{amssymb}
\usepackage{slashed} 
\usepackage{color}
\usepackage{xcolor}
\usepackage{tabularx}
\usepackage{hyperref}
\usepackage{bm}% bold math
\usepackage{cite}  %  to make successive citations hyphened
\usepackage{mathrsfs}
\usepackage{framed}

\def\abs#1{\left| #1\right|}

\colorlet{grayline}{gray!70}
\definecolor{blueline}{rgb}{0,0.27,0.55}
\definecolor{DarkGray}{gray}{0.4}
\definecolor{Gray}{gray}{0.6}
\definecolor{oucrimsonred}{rgb}{0.6, 0.0, 0.0}
\definecolor{persianblue}{rgb}{0.11, 0.22, 0.73}
\definecolor{forestgreen}{rgb}{0.13,0.35,0.13}
 \hypersetup{colorlinks, citecolor=forestgreen, linkcolor=forestgreen, urlcolor=forestgreen}
\newcommand{\be}{\begin{equation}}
\newcommand{\ee}{\end{equation}}
\newcommand{\bea}{\begin{eqnarray}}
\newcommand{\eea}{\end{eqnarray}}

\newcommand*\xbar[1]{%
  \hbox{\;%
    \vbox{%
      \hrule height 0.5pt % The actual bar
      \kern0.5ex%         % Distance between bar and symbol
      \hbox{%
        \kern-0.25em%      % Shortening on the left side
        \ensuremath{#1}%
        \kern-0.07em%      % Shortening on the right side
      }%
    }%
  }%
} 
\newcommand{\com}[1]{}
\newcommand{\gsim}{\lower.7ex\hbox{$\;\stackrel{\textstyle>}{\sim}\;$}}
\newcommand{\lsim}{\lower.7ex\hbox{$\;\stackrel{\textstyle<}{\sim}\;$}} 

\newcommand{\bc}{\begin{center}}
\newcommand{\ec}{\end{center}}

\begin{document}
\thispagestyle{empty}
\bc
{\Large {\bf 
    Can the QCD axion feed a dark energy component?}
}
\vspace*{0.4cm}
{
{\bf K. M\"u\"ursepp$^{{a}}$} 
}\\

\vspace{0.5cm}
       {\small\it
(a) Laboratory of High-Energy and Computational Physics, NICPB, R\"avala pst 10, \\ 10143 Tallinn, Estonia
\\[1mm]
}

\ec

 \vskip0.1cm
\bc
{\color{DarkGray}
\rule{0.7\textwidth}{0.5pt}}
\ec
\vskip0.5cm
\bc
{\bf ABSTRACT}
\ec
This conference thesis summarizes my presentation at Tartu Tuorla Cosmology meeting based on a collaborative work with E.Nardi and C.Smarra \cite{Muursepp:2024mbb}. It is well known that a pseudo-Nambu Goldstone boson (pNGB) coupled to a confining gauge group obtains a non-zero contribution to its mass through instanton effects. At non-zero temperature, this manifests in temperature-dependent mass $m^{2}(T) \propto T^{-n}$. If these particles come to dominate the energy density of the Universe in the non-relativistic regime they would accelerate the expansion of the Universe for $n>2$, thus providing a dark energy (DE) component. In this work, we outline a scenario in which a pNGB $\phi_{b}$ presently undergoing confinement could realize such a scenario. Using energetic considerations we find that $\phi_{b}$ alone is not enough to produce the experimentally observed amount of DE. However, coupling $\phi_{b}$ to the QCD axion $\phi_a$ allows the transfer of energy from the QCD axion to the pNGB via non-adiabatic level-crossing thus reproducing the observed amount of dark matter (DM) and DE.

\vspace*{5mm}

\noindent

%%%%%%%%%%%%%%%%%%%%%%%%%%%%%%%%%%%%%%%%%%%%%%%%%%%%%%%%%%%%%%%%%%%
\newpage
\pagestyle{plain}
%\tableofcontents

\newpage
\section{Introduction\label{sec:intro}} 

QCD axion models are presently one of the most popular extensions of the SM providing a solution to the strong-CP problem \cite{Peccei:1977hh, PhysRevLett.40.223, PhysRevLett.40.279} as well as presenting a dark matter candidate. \cite{Abbott:1982af,Dine:1982ah, Preskill:1982cy} Inspired by the QCD axion, the phenomenology of a more general class of pNGB-s, dubbed axion-like particles (ALPs) have also been extensively studied in the literature. The relative simplicity of QCD axion and ALP models means that many implications of these models for cosmology and collider physics have been considered. For recent reviews see \cite{DiLuzio:2020wdo, Marsh:2015xka, Choi:2020rgn}. In this conference thesis, we will be mainly interested in looking at the implications of the QCD axion to the DE and DM components of the Universe. \newline

The introduction of the QCD axion was originally motivated by the strong-CP problem \cite{Peccei:1977hh}, which we briefly review below. The  $SU(3)_{c}$ gauge group of the Standard Model (SM) necessitates the presence of a CP-violating term \cite{PhysRevLett.37.172, Callan:1976je }
    \begin{equation}
        \mathcal{L} \supset \theta  \Tilde{G}^{\mu \nu}G_{\mu \nu}, \text{ with } \Tilde{G}^{\mu \nu} = \epsilon^{\rho \sigma \mu \nu} G_{\rho \sigma},
    \end{equation}
with $G^{\mu \nu}$ denoting the gluon field strength tensor, that cannot be rotated away due to the non-zero masses of quarks. \cite{Cohen:1999kk, PhysRevD.92.054004, Alexandrou:2020bkd} In particular, the non-zero masses of the SM quarks provide another CP-violating parameter, meaning that the physical CP-violating parameter is given by
\begin{equation}
    \Bar{\theta} = \theta + \arg ( \det Y_u Y_d)
\end{equation}
where $Y_u, Y_d$ are the Yukawa matrices for the up and down quarks. The strong-CP-violating parameter $\Bar{\theta}$ appears explicitly in the calculation of the neutron dipole moment defined via a dimension five operator of the form:
\begin{equation}
    \mathcal{L}_{D=5} \supset  - d_{n} \frac{i}{2} \Bar{n} \sigma_{\mu \nu} \gamma_5 n F^{\mu \nu}.
\end{equation}
Comparing the strong experimental constraints placed on $d_n$ \cite{PhysRevD.92.092003}
\begin{equation}
    \abs{d^{\text{exp}}_{n}} \lesssim 10^{-26}
\end{equation}
with the results of theoretical calculations \cite{Pospelov:1999mv}
\begin{equation}
    d_{n} \simeq 10^{-16} \, \Bar{\theta} \, \text{cm}.
\end{equation}
imply a tight bound on the $\Bar{\theta}$ parameter
\begin{equation}
    \Bar{\theta} \lesssim 10^{-10}
\end{equation}
that cannot be easily explained away by anthropic arguments \cite{Ubaldi:2008nf, Dine:2018glh, DiLuzio:2020wdo}. The fine-tuning of $\Bar{\theta}$ constitutes the so-called strong-CP problem, which is most elegantly explained by the Peccei-Quinn (PQ) mechanism.  \cite{Peccei:1977hh, PhysRevLett.40.223, PhysRevLett.40.279}
\newline 

The PQ mechanism introduces a pNGB field $a$ with the Lagrangian,
\begin{equation}
        \mathcal{L}_{a} = \frac{1}{2}\left( \partial_\mu a \right)^2 + \mathcal{L} \left( \partial_{\mu} a, \phi, \psi \right) + \frac{g_s^2}{32\pi^2} \frac{a}{f}\Tilde{G}^{\mu \nu}G_{\mu \nu}.
    \end{equation}
where $f$ approximately corresponds to the scale of the breaking of the $U(1)_{PQ}$ symmetry and  $\mathcal{L} \left( \partial_{\mu} a, \phi, \psi \right)$ includes interactions of $a$ with other scalars or fermions. The coupling of the axion field to the gluons results from the $SU(3)_{c} \times U(1)_{PQ}$ anomaly and includes the explicit breaking of the shift symmetry of the theory. In more detail, under the shift symmetry $a \rightarrow a + \kappa $, the action remains invariant up to
\begin{equation}
    \delta S =  \frac{\kappa}{32 \pi^2} \int d^4 x \Tilde{G}^{\mu \nu} G_{\mu \nu}
    \label{eq: noninvariant shift symm}
\end{equation}
The breaking of the shift-symmetry by the anomaly renders the QCD axion a pNGB and thus gives it a non-zero mass. It also means that we can use the shift symmetry to remove the $\Bar{\theta}$ - term from the Lagrangian. By explicitly computing the QCD axion potential, it can then be checked that in the minimum of the axion potential, we have 
\begin{equation}
    \expval{a} = 0,
\end{equation}
thus solving the strong-CP problem. \newline

An interesting feature of axion models featuring the $\frac{a}{f} G \Tilde{G}$ coupling is that the mass generated by the axion-gluon interaction is changing with temperature. In particular, the mass $m_a$ generated by the axion-gluon interaction is determined by correlation functions of the gluon fields and is related to the previously defined energy scale f as follows: 
\begin{equation}
    f^2 m_a^2 = i \int \text{d}^{4}x\expval{ \frac{\alpha_s}{8 \pi} G^{\mu \nu} \Tilde{G}_{\mu \nu}(x) \frac{\alpha_s}{8 \pi} G^{\alpha \beta}\Tilde{G}_{\alpha \beta} (0)} \equiv \chi
\end{equation}
where $\alpha_s$ denotes the fine structure constant of $SU(3)_{c}$ and $\chi$ is known as the topological susceptibility. Importantly, $\chi$ is not a constant function and depends on the temperature of the ambient plasma in the Universe. To understand this behaviour, let us first analyse the case of the QCD axion and then apply the same arguments to the case of a general pNGB coupling to the gauge bosons of a confining group.\newline

At temperatures well above the QCD confining temperature $T_c$, the axion 
is massless $\chi(T) = 0$ due to screening of the free color charges in the plasma. As $T$ decreases  towards $T_c$,  
color charges get confined into color singlets, and a mass is generated, 
which  reaches its zero-temperature  value $m_a \sim \Lambda_{QCD}^2/F$ at $T\lsim T_c$ 
where $\Lambda_{QCD} \sim 100 \text{ MeV} $ is the QCD scale, and 
$F\gtrsim 10^9\,$GeV is the axion decay constant. 
More generally, the dependence of the axion mass on temperature can be parameterised as $ m^2_a(T) \sim T^{-n}$. The dilute instanton gas approximation
(DIGA)~\cite{Callan:1977gz,Callan:1978bm}  at the lowest order predicts that
$n=\beta_0 +n_f-4$ where $n_f$ is the number of light quarks 
and $\beta_0 = \frac{11}{3}N - \frac{1}{3} n_s T_s -\frac{4}{3} n_f T_f $ 
is the one loop coefficient of the $\beta$-function,  $N$ is the degree of the confining gauge group,
 $n_s$ the number of light scalars and $T_{s,f}$ the index of the 
corresponding representations ($T_{s,f}=1/2$ for the fundamental). Applying this to the case of QCD with $n_f=3,(0)$ 
one obtains $n=8,(7)$. \newline

However, the simple paramerisation introduce above thus not give the full picture, for the index $n$ introduced above is itself also a function of temperature $n = n(T)$. The precise dependence of $n$ on temperature has been investigated in lattice simulations and it has been found that while n remains close to DIGA results at temperatures $T >> T_{c}$, instead for temperatures close to the oscillation of the axion field, the dependence of the axion mass on temperature is significantly milder than the DIGA prediction. \cite{Lombardo:2020bvn} \newline

In particular, the case $n=6$ is particularly interesting in models with a pNGB $\phi_{b}$ coupled to a dark gauge group $G_{b}$ that is presently undergoing confinement. \cite{Muursepp:2024mbb}. This is because in such case the corresponding masss is still increasing with decreasing temperature $m_{b}^2 \sim T^{-n}$, which in the non-relativistic regime could thus behave like DE. In more detail, the conservation law resulting from the continuity equation
\begin{equation}
    d(\rho_b a^3) = -p_b da^3
\end{equation}
can be used to determine an effective equation of state
\begin{equation}
    p_b = w \rho_b,
\end{equation}
where $\rho_b$ denotes the energy density of $\phi_b$ , $p_b$ denotes the pressure and $a$ the scale factor. For non-relativistic populations of $\phi_b$, we have $\rho_b = m_b (T) n_{b}$ where $n_b$ is the number density of $\phi_b$, implying $w = -n/6$. It immediately follows that if $\phi_b$ starts to dominate the energy density of the Universe, it has an accelerating effect already for very mild temperature-dependence of $m_b$, corresponding to $n > 2$ and for $n=6$ it behaves like a cosmological constant. \newline 

Unfortunately, this mechanism cannot work for a single pNGB unless additional energy is injected into the system. This is because, by simple energetic considerations, the energy density of a single axion presently undergoing confinement falls below the energy density of radiation which is itself much smaller than the energy density of dark energy today. \newline 

Alternatively, one can consider a two-axion system featuring a \textit{level-crossing} mechanism that allows the transfer of energy between the QCD axion $\phi_a$ and the pNGB $\phi_b$. The level-crossing mechanism has been studied before in the adiabatic regime in the context of dark matter abundance, domain wall formation, and isocurvature perturbations.\cite{Kitajima:2014xla, PhysRevD.92.063512, PhysRevD.93.075027, Ho:2018qur, Cyncynates:2023esj,Li:2023uvt} In contrast to those studies, our work focuses on the non-adiabatic regime involving only the conversion of a small fraction of $\phi_a$ number density into the number density of $\phi_b$ and allows to reproduce both the observed amount of DM and DE.\newline

\section{Two axion system}

To make things more concrete, let us consider two non-abelian gauge groups $G_a \times G_b$ that confine at energies $\Lambda_b$ and $\Lambda_a$ with $\Lambda_b << \Lambda_a$. Moreover, we introduce two complex scalars $\Phi_1$ and $\Phi_2$, and two fermions $\psi$ and $\chi$ that transform under  $G_a \times G_b$ as $(1,3)$ and $(3,3)$ respectively. The instantons of the two gauge groups are responsible to generating the potential for the two-axion system, which takes the form
\begin{equation}
    V(\phi_a, \phi_b) = \Lambda_a^4 \left[ 1- \cos \left( \frac{\phi_a}{F}\right) \right] + \Lambda_b^4 \left[ 1 - \cos \left( \frac{\phi_a}{F'} + \frac{\phi_b}{f} \right) \right]
\end{equation}
where $F, F \propto v_1$ and $f \propto v_1$ where $v_1,v_2$ are the vacuum expectation values of the radial modes of $\Phi_1$ and $\Phi_2$ and the fields $\phi_a$ and $\phi_b$ are defined in terms of linear combinations of $a_1 = \arg(\Phi_1)$ and $a_2 = \arg(\Phi_2)$ :
\begin{equation}
    \begin{pmatrix}
    \phi_a \\ \phi_b
    \end{pmatrix} = \begin{pmatrix}
        \cos \beta & \sin \beta \\
        -\sin \beta & \cos \beta
    \end{pmatrix}
    \begin{pmatrix}
        a_1 \\
        a_2
    \end{pmatrix}
\end{equation}
The equations of motion of the two-axion system behave like a coupled forced harmonic oscillator with the damping force controlled by the expansion of the Universe. In particular:
    \begin{equation}
        \ddot{A} + 3H \dot{A} + M^2 A = 0
    \end{equation}
    where 
    \begin{equation}
    \begin{split}
       &A = \begin{pmatrix}
           \phi_a \\ \phi_b \end{pmatrix} \hspace{0.3cm} M^2 =m_a^2  \begin{pmatrix} 1 & \epsilon r(T) \\  \epsilon r(T) & r(T) \end{pmatrix} \hspace{0.3cm} \\ &m_a = \frac{\Lambda_a^2}{F} \hspace{0.3cm} r(T) = \frac{m_b^2(T)}{m_a^2} \hspace{0.3cm} \epsilon \simeq \frac{f}{F}
           \hspace{0.3cm} 
    \end{split}
    \end{equation}
Associating $\phi_a$ with the QCD axion sets $\Lambda_a \simeq 160 \text{ MeV} $ and requiring that the "dark" axion $\phi_b$ is presently undergoing confinement sets $\Lambda_b \lesssim 10^{-4}$. When the $SU(3)_{a}$ confines the QCD axion obtains its zero temperature mass $m_a^2 = \frac{\Lambda_a^2}{F}$, while the second axion $\phi_b$ is still massless at such high temperatures due to the screening of the unconfined particles that are charged under $SU(3)_b$ only. Thus we have $m_a >> m_b$ for $T >> T_b$, where $T_b$ denotes the temperature at which $SU(3)_b$ confines. The zero temperature mass of $\phi_b$ is given by $m_b = \frac{\Lambda_b^2}{f}$ so that choosing $f << F$, we can arrange that $m_a < m_b$ close to the confinement of $SU(3)_b$. \newline 

\begin{figure}
    \centering
\begin{tabular}{ll}

\includegraphics[width=.35\linewidth]{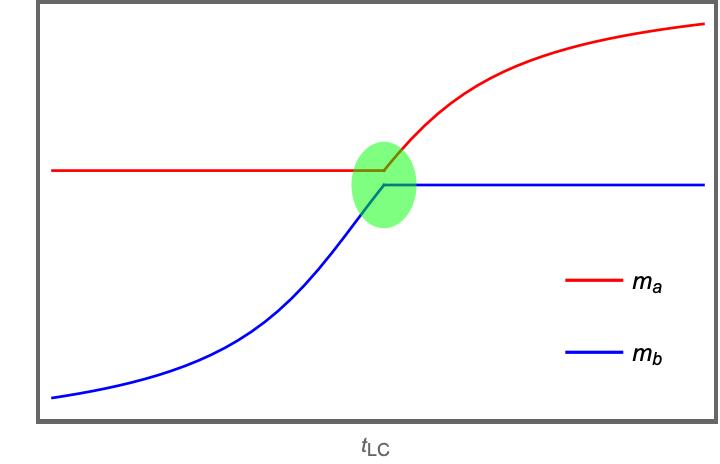} & \includegraphics[width=.35\linewidth]{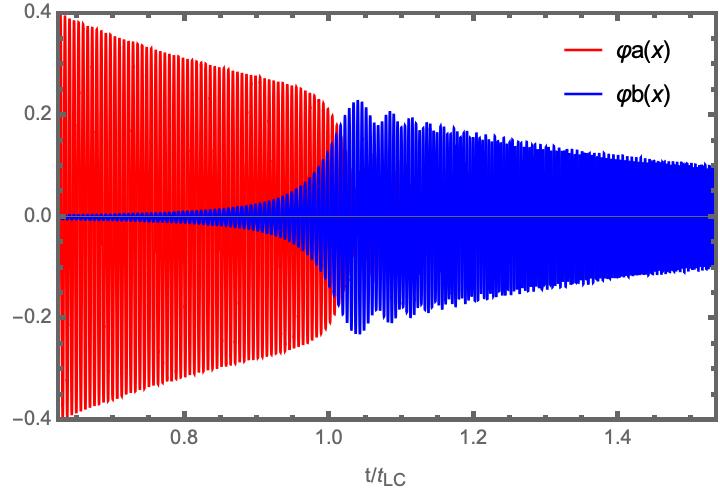} \\
\includegraphics[width=.35\linewidth]{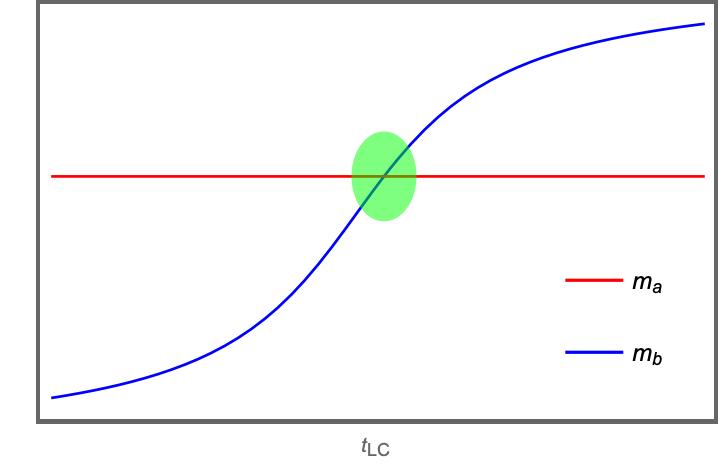} & \includegraphics[width=.35\linewidth]{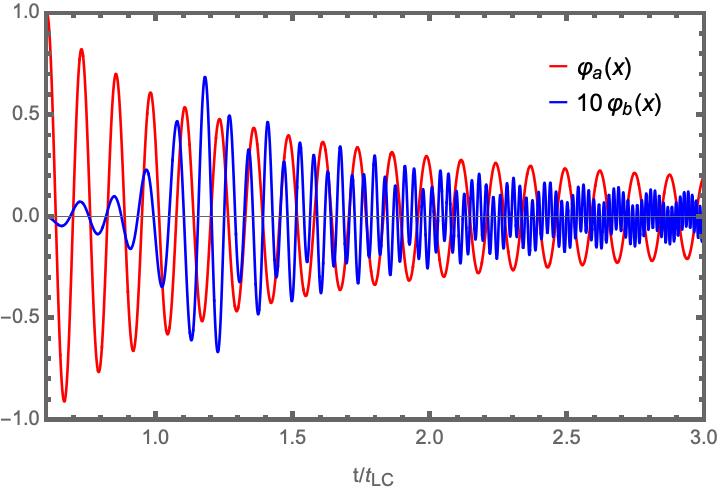} 
\end{tabular}  
\caption{Difference between adiabatic ($m_a (\epsilon t_{LC}) >> 1$) (upper row) and non-adiabatic $m_a (\epsilon t_{LC}) < 1$)  (lower row) level crossing. Pictures made by E.Nardi. }
\label{fig: adiabatic and diabatic LC}
\end{figure}

Thus we choose parameters such that $m_a > m_b$ at early times, and $m_b < m_a$ at late times. Therefore there must be a region where $m_b(T_{LC}) \simeq m_a$ which defines the level crossing temperature $T_{LC}$. The level crossing phenomenon in the context of axions has commonly been studied in the \textbf{adiabatic} regime, which is realized when many axion oscillations occur during level crossing. Since the typical duration of level crossing is given by $\Delta t_{LC} \simeq \epsilon t_{LC}$ where $t_{LC}$ can be found from $T_{LC}$ using the relation between time and temperature in matter domination era, and the period of one oscillation
\footnote{It is the same for both axions around level crossing since around that time $m_a \simeq m_b(t)$.} is roughly given by $t_{osc} = m_a^{-1}$ the adiabaticity condition can be expressed as
\begin{equation}
    m_a \epsilon t_{LC} >> 1.
\end{equation}
If this condition is satisfied,  the heavier and lighter eigenstates maintain their identities after level crossing and thus energy is adiabatically transferred from $\phi_a$ to $\phi_b$.  Instead, the opposite scenario,
\begin{equation}
    m_a \epsilon t_{LC} \lesssim 1.
\end{equation}
results in either \textbf{non-adiabatic} (or also known as \textbf{diabatic}) level crossing where only tiny amount of energy is transferred between the axions or no energy is transferred at all. These two cases are  visualized on Figure \ref{fig: adiabatic and diabatic LC}

\section{Dark energy from non-adiabatic level crossing}

Having discussed the general properties of the two-axion system let us now come back to the QCD axion and dark axion scenario. This scenario introduces several constraints between $f,F, \Lambda_a, \Lambda_b$ that when combined appear to only be compatible with non-adiabatic conversion. \newline 

In order for the $\phi_b$ axion to be still undergoing confinement today we require $\Lambda_b < T_0$, where $T_0$ is the temperature of the CMb today. Moreover, to match the observed amount of DE by matter-dark energy equality (defined by the temperature $T_{DE}$) we also require that  $T_{DE}< T_{LC}$. Before, we have seen that to realize the level crossing mechanism we require $f << F$, however to ensure that the $U(1)$ symmetry associated with the pNGB $\phi_b$ is broken by the time of level crossing, we must also ensure that $T_{LC} < f$. \newline

Recalling that $\Lambda_b \simeq 10^{-4} \text{ eV}$ we can then choose $f \gtrsim 10^{-2} \text{ eV}$ which implies a pre-inflationary scenario with $F \gtrsim 10^{14} \text{ eV}$ and $m_a \lesssim 6 \times 10^{-8} \text{ eV}$. In order to reproduce the observed amount of dark matter only mild tuning of the misalignment angle is needed: $\theta_a \lesssim 6 \%$. For typical value of level crossing time $t_{LC} \simeq 10^{9}$ years, we thus have 
\begin{equation}
    m_a t_{LC} \epsilon \simeq 1,
\end{equation}
requiring that the level crossing happen in the non-adiabatic regime. As a consistency check, we can compute the ratio of DE and DM at $t_{LC}$ which can be interpreted as the energy density converted from the QCD axion to the dark axion providing dark energy. We have
\begin{equation}
    \frac{\rho_{DE}}{\rho_{DM}} \bigg \rvert_{t_{LC}} =
\left(\frac{1+z_{DE}}{1+z_{LC}}\right)^3\sim 1\%-2\%,
\label{eq: consistency}
\end{equation}
where $z_{DE} \sim 0.5$ is the redshift at dark energy-dark matter equality and $z \sim 5$ the redshift at the level crossing. Equation \eqref{eq: consistency} thus shows that only a few percent of QCD axions should be converted to the dark axions providing DE, consistently with our chosen parameters.

\begin{figure}
    \centering
        \includegraphics[width= 0.7\textwidth]{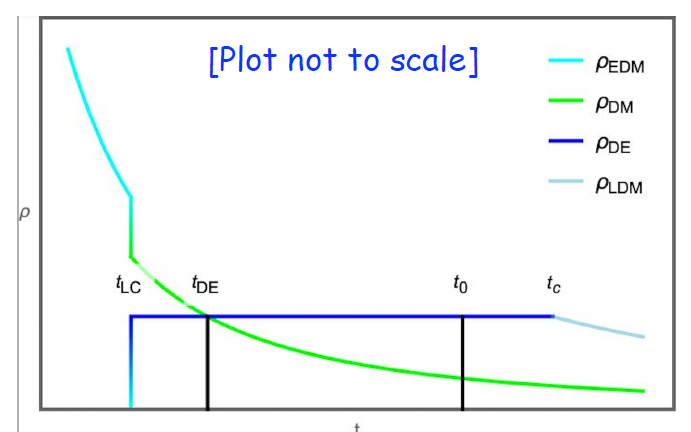}
        \caption{The cosmological evolution of dark energy and dark matter in the level crossing scenario. About $1-2\%$ needs to be converted from the DM to DE to explain the energy budget of the Universe. Picture made by E.Nardi.}
        \label{fig:illustration}
\end{figure}

Finally, for illustration purposes in figure \ref{fig:illustration} we show the cosmological evolution of the two-axion system. The light blue line $\rho_{EDM}$ denotes the energy density of the early dark matter component sourced by the QCD axion. At that time the contribution of $\phi_b$ to the energy density is negligible. At level crossing a small fraction of $\phi_a$ is converted to $\phi_b$ introducing a small drop in energy density of DM (now denoted by $\rho_{DM}$) while also sourcing the production of a substantial amount of dark energy $\rho_{DE}$. After $t_{LC}$ the DM and DE components scale as $\rho_{DE} = \text{const.}$ and $\rho_{DM} \propto a^{-3}$ where $a$ denotes the scale factor. Finally at a later time $t_c$ when $SU(3)_{b}$ confines, the mass of $\phi_b$ obtains a constant value and so $\phi_{b}$ starts to behave like late dark matter ($\rho_{DM}$) with a substantially larger energy density than in the present Universe.

\section{Conclusion and discussion}

In this conference thesis, we presented a model of two-axions that can explain both the observed amount of DE and DM in the Universe. We focused mainly on the case of a QCD axion coupled to a pNGB with varying mass, still undergoing confinement today. Due to the huge hierarchy between the confinement scales of the QCD axion and the  pNGB, we found that the energy transfer between the two axions can only happen through non-adiabatic conversion and that this scenario is highly constrained. The hierarchy of scales also complicates numerical studies due to extremely fast oscillations of the axions at late times and the short duration of the level-crossing, Thus, the presently considered two-axion system can only be studied in detail with more advanced numerical methods which is why we have limited ourselves to intuitive estimates focusing on the viability of the mechanism. Alternatively, one could focus on a scenario including two pNGB-s unrelated to the QCD axion. This would allow to reduce the confinement scale $\Lambda_a$, potentially enabling to convert more of $\phi_a$ into $\phi_b$ and making numerical studies more tractable. Moreover, giving up on the QCD axion may also make the construction less constrained. 

\section*{Acknowledgements}

While working on this project I was supported  by the 
Estonian Research Council grant PRG803.
We acknowledge support from the CoE grant TK202 “Foundations of the Universe” and from the 
CERN and ESA Science Consortium of Estonia, grants RVTT3 and RVTT7.   
This article is based in part upon work from COST Action COSMIC WISPers CA21106, supported 
by COST (European Cooperation in Science and Technology).
We thank the Galileo Galilei Institute for Theoretical Physics, where this work was started, for hospitality.

%%%%%%%%%%%%%%%%%%%%%%%%%%%%%%%%%%%%%%%%%%%%%%%%%%%%%%%%%%%%%%%%%%%%%%%%%%

\bibliographystyle{jhep}
\bibliography{axionthesis}

%%%%%%%%%%%%%%%%%%%%%%%%%%%%%%%%%%%%%%%%%%%%%%

\end{document}